\documentstyle[12pt,epsfig]{ioplppt}

\begin{document}
\jl{4}

\letter{Coherent radiation of an ultra-relativistic charged particle 
channeled in a periodically bent crystal
\footnote{published in
J. Phys. G: Nucl. Part. Phys. {\bf 24} (1998) L45--L53, Copyright 1998
IOP Publishing Ltd., http://www.iop.org
}
}

\author{Andrei V Korol\dag\ftnote{4}{E-mail: korol@rpro.ioffe.rssi.ru}, 
Andrey V Solov'yov\ddag\ftnote{3}{E-mail: solovyov@rpro.ioffe.rssi.ru},  
and  Walter Greiner \P}

\address{\dag Department of Physics,
St.Petersburg State Maritime Technical University,
Leninskii prospect 101, St. Petersburg 198262, Russia}
\address{\ddag A.F.Ioffe Physical-Technical Institute of the Academy
of Sciences of Russia, Polytechnicheskaya 26, St. Petersburg 194021,
 Russia}
\address{\P Institut f\"{u}r Theoretische Physik der Johann Wolfgang
Goethe-Universit\"{a}t, 60054 Frankfurt am Main, Germany}

\begin{abstract}
We suggest a new type of the undulator radiation which is generated by
an ultra-relativistic particle channeled along a periodically bent
crystallographic plane or axis.  The electromagnetic radiation arises
mainly due to the bending of the particle's trajectory, which follows
the shape of the channel.  The parameters of this undulator, which
totally define the spectrum and the angular distribution of the
radiation (both spontaneous and stimulated), depend on the type of the
crystal and the crystallographic plane (axis), on the type of a
projectile and its energy, and on the shape of the bent channel, and,
thus, can be varied significantly by varying these characteristics.

As an example, we consider the acoustically induced radiation (AIR)
which is generated by ultra-relativistic particles channeled in a
crystal which is bent by a transverse acoustic wave.  The AIR
mechanism allows to make the undulator with the main parameters
varying in wide ranges, which are inaccessible in the undulators based
on the motion of particles in the periodic magnetic fields and also in
the field of the laser radiation.  The intensity of AIR can be easily
made larger than the intensity of the radiation in a linear crystal
and can be varied in a wide range by varying the frequency and the
amplitude of the acoustic wave in the crystal.

A possibility to generate stimulated emission of high-energy photons
(in kev--MeV region) is also discussed.

\end{abstract}
% insert suggested PACS numbers in braces on next line
\pacs{41.60, 61.85+p}

We suggest a new type of the undulator radiation generated by an
ultra-relativistic particle during its channeling in periodically bent
crystal.

The novel feature, introduced into the radiative channeling process by
a periodically bent channel is that the oscillations of the particle
in the transverse direction become the effective source of radiation
of an undulator type due to the constructive interference of the
photons emitted from spatially separated but similar parts of the
projectile's trajectory.
 
The phenomenon of the channeling radiation of a charged projectile in
a linear crystal (see eg. \cite{Baier}, \cite{Kniga}) as well as in a
``simple'' (i.e. non-periodic) bent channel \cite{Solov} are known,
although in the latter case the theoretical and experimental data are
much less extensive so far.  The theory and various practical
implementations of the electromagnetic radiation by a charge moving in
spatially periodic static magnetic fields (a magnetic undulator), in
the laser field (a laser-based undulator) have long history
\cite{Ginz,Motz} and are well elaborated \cite{Baier}, \cite{Kniga},
\cite{Alferov}.  It is also known (see eg. \cite{Baier}) that a
positively charged ultra-relativistic particle, undergoing a planar
channeling in a linear crystal, radiates electromagnetic waves whose
spectral and angular distributions are those of the undulator (a
natural undulator) due to the transverse oscillations caused by the
action of the repulsive interplanar potential. Although, the features
of the natural undulator radiation are blurred due to the distribution
of the particles in the beam over the transverse energy and over the
entering angle.

In this work we demonstrate, that the system
``ultra-relativistic charged particle + periodically bent crystal
channel'' represents by itself a new type of the undulator and,
consequently, serves as a new source of the undulator radiation of
high intensity levels, monochromaticity and a particular
pattern of the angular-frequency distribution.  The electromagnetic
radiation in this undulator arises mainly due to the bending of
particle's trajectory, which follows the shape of the channel.  The
parameters of this undulator, as well as the characteristics of the
electromagnetic radiation, depend on the type of the crystal and its
crystallographic plane (axis), on the type of a projectile and its
energy, and are also dependent on the shape of the bent channel, and,
thus, can be varied significantly by varying the enlisted
characteristics.

For the undulator described above the essential point is the existence
of the bent channel of a periodic shape.  Two realistic ways of a
``preparation'' of such a channel could be discussed.  It is feasible,
by means of mordern technology \cite{Biryukov}, to grow the crystal
with its channels being statically bent according to a particular
pattern.  Another possibility, a dynamically bent crystal, arises if
one considers propagation of the acoustic wave (AW) along some
particular direction in a crystal, see figure 1.  Under the action of
the AW the crystal channels, linear initially, will also be bent
periodically.  In both cases the passage of an ultra-relativistic
particle along the bent channel gives rise to the undulator radiation
due to the curvature of the trajectory, provided that the projectile
is trapped into the channel.  The latter condition is subject to the
general condition for the channeling process in a bent crystal
\cite{Solov,Biryukov}, and can be fulfilled by a proper choice of
projectile energy and maximal curvature of the channel, as described
below.

%%% Fig.1 
\begin{figure}
\hspace{2.5cm}\epsfig{file=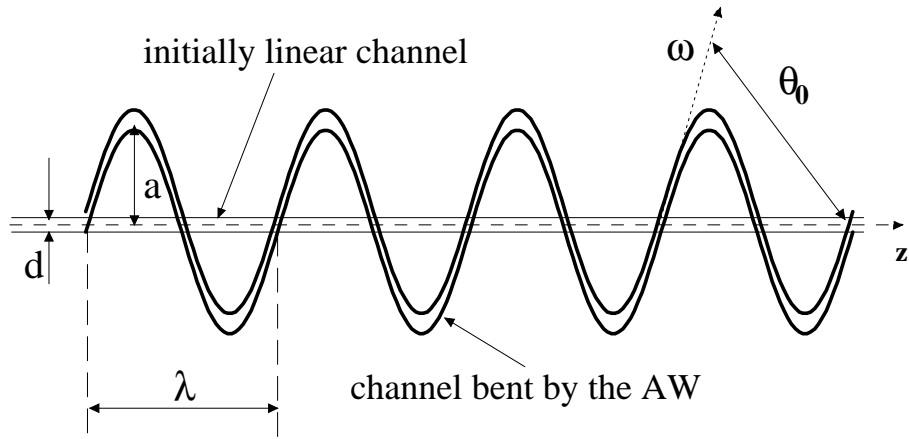,width=12cm}
\caption{Schematic representation of the initially linear planar
channel bent by the AW. The notations are: $d$ is the channel
width, $a, \lambda$ are the AW amplitude and wavelength, respectively.
The parameter $\theta_0$ is the maximal polar angle of the photon emission
(see explanations in the text).}
\end{figure}

The advantage of the static channel is that its parameters are fixed
and thus the projectile moves along the fixed trajectory as well. To
calculate the characteristics of the emitted radiation one needs to
know only the number of the periods and the local cuvature radius. The
disadvantage is that when fixing the number of periods then the
undulator parameters can be varied only by changing the energy of the
particle. This makes the undulator less tunable.

The important feature of the dynamically bent crystal by means of an
AW is that it allows to consider an undulator with the
parameters\footnote{$N_{\rm u}$ is the number of periods in the
undulator and $p=2\pi\gamma\, a/\lambda$ is its parameter (see figure
1).}  $N_{\rm u}$ and $p$ varying in wide ranges, which are determined
not only by the projectile's energy but by the AW frequency and
amplitude as well.  The two latter quantities can be easily tuned
resulting in the possibility to vary significantly the intensity and
the shape of the angular distribution of the radiation, which
hereforth will be referred as acoustically induced radiation, AIR.

The intensity of AIR can be made larger than the radiation intensity 
emitted by a charged particle channeled in an oriented linear crystal
\cite{Korol} (see also \cite{Solov}). The AIR mechanism allows to make
an undulator with the parameters $N_{\rm u}$ and $p$ changing in a
wide range which is inaccessible for undulators based on the motion of
charged particles in periodic magnetic fields and also in the
field of the laser radiation \cite{Kniga}, \cite{Alferov}.  In the
suggested scheme, AIR is generated by the relativistic charged
particles, with relativistic factors $\gamma=\varepsilon/mc^2\gg 1$
($c$ is the velocity of light, $m$ is the mass of the projectile and
$\varepsilon$ is its energy).  The large range of $\gamma$ available
in modern colliders at present or in near future for various charged
particles, both light and heavy, and the wide range of frequencies and
amplitudes possible for AW in crystals both allow to generate the AIR
photons with the energies up to the TeV region \cite{Korol}.

The specific pattern of the undulator radiation combined with the AIR
mechanism allows, to our opinion, to discuss the possibility of creating
a powerful source of stimulated monochromatic radiation of high-energy
photons.

In this letter we consider the simplest case, when the initially
linear planar channel (the case of the axial channel is treated
analogously) is bent periodically under the action of the transverse
monochromatic AW transmitted along the channel (see figure 1) with the
velocity $v_{\rm t}$.  Let $\nu$, $\lambda=v_{\rm t}/\nu$ and $a$ be
the AW frequency, wave length and amplitude, respectively.  The
bending of the channel becomes significant if the AW amplitude, $a$, is
larger than the interplanar distance, $d$, i.e.\ $a\gg d$.

The conditions which must be fulfilled to consider the system ``AW +
channeled particle'' as an undulator are the following.  Let the time
of flight, $\tau =L/c$, of the particle through the crystal of the
thickness $L$, be much smaller than the AW period $T$: $\tau \ll
T$. This allows to disregard the variations of the shape of the
channel during the time $\tau$.

The channeling process in a bent crystal takes place if the maximal
centrifugal force in the channel, $m \gamma v^2/R_{\rm min}$ (where
$R_{\rm min}$ is the minimum curvature radius of the bent channel) is
less than the maximal interplanar field $U_{\rm max}^{\prime}$
\cite{Solov, Biryukov}:
\begin{equation}
m \gamma v^2/R_{\rm min} < q\, U_{\rm max}^{\prime}.
\label{1}
\end{equation}
where $q$ is the charge of the projectile.  Provided (\ref{1}) is
fulfilled, the projectile, which enters the crystal under the angle
$\theta$ much less than the critical angle $\theta_{L}$, will move,
being trapped in the channel, along the trajectory:
\begin{equation}
y(x)=a\sin\left(2\pi{x \over \lambda }\right),
\qquad x=[0\dots L]
\label{2}
\end{equation}
The minimum curvature radius of this trajectory is equal to $R_{\rm
min}=(\lambda/2\pi)^2/a$.  Thus, a decrease in $R_{\rm min}$ and,
consequently, an increase in the maximum acceleration of the particle
in the channel is achieved by decreasing $\lambda$ and increasing $a$.
As a result, photon emission due to the projectile's acceleration in
the bent channel may be significantly enhanced.  This radiation is
emitted coherently from similar parts of the trajectory, and may
dominate \cite{Korol} over the radiation caused by the acceleration of
the particle in the linear channel.

Both the motion of the projectile in the bent channel and the spectrum
of the generated electromagnetic radiation are of the undulator-type,
only if $\lambda \ll L$, i.e. if the channeling particle oscillates
many times within the length $L$ of the crystal.  As any other
undulator \cite{Baier}, the suggested undulator is characterized by
the undulator frequency, $\omega_0$, and the undulator parameter, $p$.
These quantities, which determine the radiation spectrum, are
dependent on $\lambda$, $a$ and on the relativistic factor $\gamma$,
and are defined as follows:
\begin{equation}
\omega_0 = 2\pi\, {c \over \lambda}, \qquad
p = 2\pi \,\gamma \, {a \over \lambda}
\label{3}
\end{equation}
For the trajectory (\ref{2}) the relation (\ref{1}) reads as
\begin{equation}
\nu^2\, a < C \equiv \gamma^{-1}\cdot
\left({ v_{\rm t} \over 2\pi} \right)^2 \cdot
\left({ q\, U_{\rm max}^{\prime} \over mc^2}\right)
\label{4}
\end{equation}
\noindent
and determines the ranges of $\nu=v_{\rm t}/\lambda$ ($\nu$ is the AW
frequency), $a$ and $\gamma$ for which the channeling process as well
as the undulator radiation, can occur for given crystal and
crystallographic plane (the parameters $U_{\rm max}^{\prime}$ and
$v_{\rm t}$ are subject to the choice of a particular crystal, a plane
or an axis) and for given projectile type, characterized by the rest
mass $m$ and the charge $q$.

Figure 2 illustrates the ranges of $\nu$ (in Hz) and $a$ (in cm) in
which the channeling process is possible for a positron and a proton
both of energy $50\ GeV$ in a carbon crystal near the $(110)$
crystallographic plane.  The solid thick line represents the boundary
$\nu^2\, a = C$, so that the range of validity of (\ref{4}) lies below
this line.  We used the following values: $v_t = 11.6\cdot10^5$ cm/s
\cite{Mason} and $U_{\rm max}^{\prime}= 12.0$ Gev/cm \cite{Baier}.
The dotted and the dashed-dotted lines indicate the constant values of
the undulator parameter $p$.  The dashed lines correspond the constant
values of the parameter $N$ both for positron and proton, which is
defined as the number of the AW periods per 1 cm: $N\, ({\rm cm^{-1}})
= \lambda^{-1}$.  Figure 2 demonstrates that the parameters $p$ and
$N$ vary in wide ranges: $N=1...100$, $p=0.1-500$ for projectile
positron and $p= 0.001-0.1$ for proton.  The upper limiting values of
$p$'s are larger by more than an order of magnitude than those
accessible in the undulators based on the motion of the charged
particles in periodic magnetic fields and also those in the field of
the laser radiation \cite{Kniga}, \cite{Alferov}.

%%% Fig.2
\begin{figure}
\hspace{2.5cm}\epsfig{file=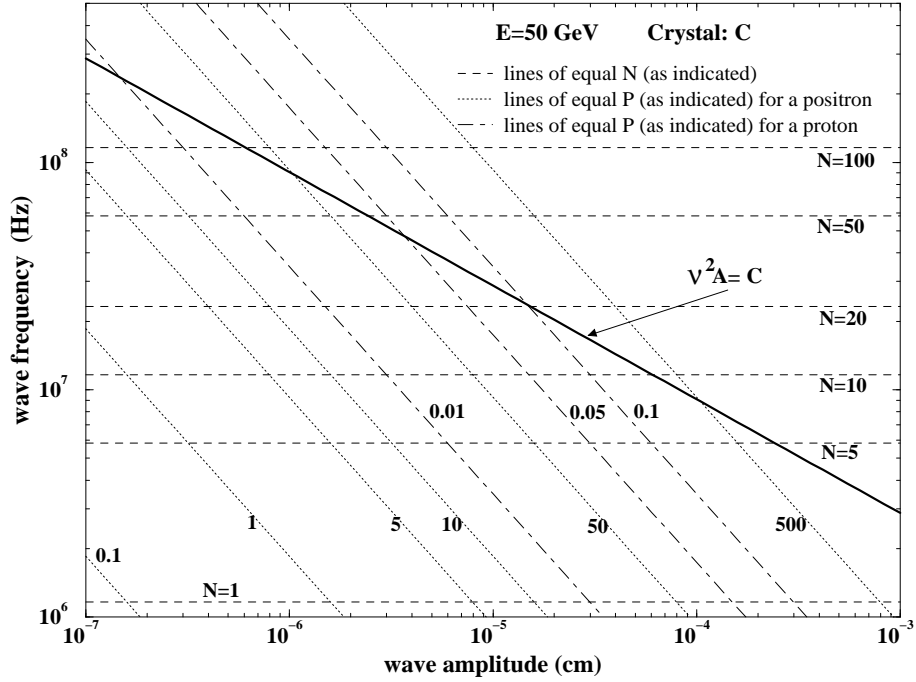,angle=270,width=12cm}
\caption{The ranges of $\nu$ (in Hz) and $a$ (in cm) in which the
channeling process is possible for 50 GeV positron and proton in a
carbon crystal along the (110) plane.  See also the explanations in
the text.}
\end{figure}

To obtain the characteristics of the radiation, the angular and the
spectral distributions, in the case of an ultra-relativistic channeled
particle, one may use the quasi-classical approach \cite{Baier}.  In
the limit $N_{\rm u},\, p^2 \gg 1$, achievable for a projectile
positron (see figure 2), the angular and the spectral distribution of
the radiated energy can be represented in the form:
\begin{eqnarray}
\fl {\d E \over \hbar\, \d\omega\, \d \Omega }
= 16 \alpha \, {N_{\rm u}^2 \gamma^2 \over p^2}\, {D_N(\eta) \over 1+u
}\,
\left({4x^2 \over \mu }\right)^{2/3}\,
\Biggl\{
\biggl[
{\theta^2 \sin^2\varphi \over 2\, \theta_0^2} +
\Delta\, z\, \left({\mu \over 4x^2}\right)^{1/3} 
\, \biggr]\, \cos^2\psi\, {\rm Ai}^2(z) 
\nonumber \\
\lo{+}
(1+\Delta)\, \left({\mu \over 4x^2}\right)^{1/3}\, 
\sin^2\psi\, {\rm Ai}^{\prime\,2 }(z)  
\Biggr\}
\label{5}
\end{eqnarray}
where ${\rm Ai}(z),{\rm Ai}^{\prime}(z)$ are the Airy function and its
derivative respectively, $\alpha$ is the fine structure constant,
$\omega$ is the photon frequency, $\theta$ and $\varphi$ are the polar
and the azimuthal angles of the photon emission.  The $z$ axis is
chosen along the direction of the AW transmission.  The parameter
$\theta_0$, related to $p$ and $\gamma$ through $\theta_0^2 =
p^2/(2\gamma^2)$, defines the maximal (with the accuracy of
$\gamma^{-1}$) polar angle of the photon emission, as illustrated in
figure 1.  The asymptotic expression (\ref{5}) is valid for $\theta <
\theta_0$.

Other notations used are as follows:
\numparts
\begin{eqnarray}
\fl D_N(\eta) = \left({\sin N_{\rm u}\pi\eta \over N_{\rm u}\,
\sin \pi\eta}\right)^2,
\qquad
\eta = x\, \left[1 + {2\over p^2} + {\theta^2 \over \theta_0^2}\right],
\qquad x = {\omega^{\prime} \over \omega_0}\, {p^2 \over 4\gamma^2}
\label{6a} \\
\fl z = \left({4x^2 \over \mu}\right)^{1/3}\left[{1\over p^2} +
{\theta^2 \sin^2\varphi \over 2\theta_0^2} \right],
\qquad \mu = 1 - \nu^2,
\qquad \nu =  {\theta \cos\varphi \over \sqrt{2}\, \theta_0},
\label{6b} \\
\fl \psi = \eta\left( {\pi \over 2} - {\rm arcsin}\,\nu\right)
-3x\,\nu\,\mu^{1/2},
\quad 
\Delta = {u^2 \over 2(1+u)},\quad 
u = {\hbar \omega \over \varepsilon -\hbar \omega  }
\quad
\omega^{\prime}
=\omega\,(1+ u)
\label{6c}
\end{eqnarray}
\endnumparts
The parameter $u$ takes into account the correction due to the
radiative recoil \cite{Baier}. In the classical limit ($\hbar=0$)
$u=0$ and $\omega^{\prime}=\omega$.

Expression (\ref{5}) together with the definitons
(\ref{6a})-(\ref{6c}) clearly exhibits the features intrinsic to the
undulator radiation \cite{Alferov}. The intensity of radiation is
proportional to $N_{\rm u}^2$ ($N_{\rm u}=L/\lambda$).  The spectral
and angular behaviour of the radiation are determined mainly by the
function $D_N(\eta)$, which has sharp maxima (each of constant width
$2/N_{\rm u}$) at the points $\eta=K=1,2,3\dots$. Combined with the
definitions of $\eta$ and $x$ from (\ref{6a}) it defines the set of
harmonics, $\omega^{\prime}_K$, emitted at the angle $\theta$, as well
as the width of each harmonic, $\Delta \omega^{\prime}$, which is
independent on $K$:
\begin{equation} 
\omega^{\prime}_K = {4\gamma^2 \omega_0\, K 
\over p^2 \left(1 + {2\over p^2} + {\theta^2 \over \theta_0^2}\right)},
\qquad
\Delta \omega^{\prime} = {2 \over N_{\rm u}}\, {\omega^{\prime}_K \over K} 
\qquad K=1,2,3\dots
\label{7}
\end{equation}

The Airy function and its derivative, both satisfying the conditions
${\rm Ai}(z),|{\rm Ai}^{\prime}(z)|\sim 1,\ z\le 1$ and ${\rm
Ai}(z),|{\rm Ai}^{\prime}(z)|\longrightarrow 0,\ z\gg 1$, define the
frequency of the radiated intensity maximum $\omega_{\rm
max}^{\prime}$.  For $\omega^{\prime}\gg \omega_{\rm max}^{\prime}$
the intensity of radiation rapidly decreases. It can be shown that
$\omega_{\rm max}^{\prime} \sim p\gamma^2\,\omega_0$.

All these features are illustrated in figures 3 and 4, where the
angular-spectral distribution (\ref{5}) is presented for a $50$ GeV
positron in a carbon crystal of the thickness $L=1$ cm.

The parameters of the AW are indicated in figure 3, which represents
the spectral distribution of the radiation in the case of the forward
emission ($\theta=0^{\circ}$).  It is seen that all harmonics are well
separated: the width of each peak is $\le 1$ MeV whilst the distance
between two neighbouring peaks is $\approx 7$ MeV.

%%% Fig.3
\begin{figure}
\hspace{2.5cm}\epsfig{file=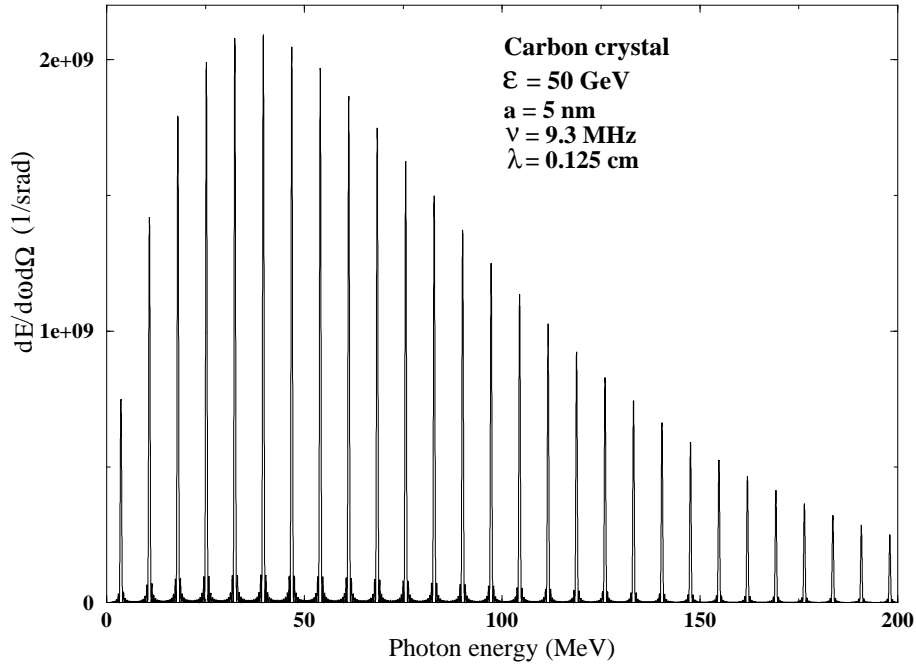,angle=270,width=12cm}
\caption{The spectral dependence of the radiated energy for 50 GeV
positron in carbon crystal. The polar angle of the emission
is $\theta=0^{\circ}$.
Other parameters as indicated.}
\end{figure}

Two 3D-plots in figures 4a (the azimuthal angle, with respect to the
undulator plane, is equal to $\varphi=0^{\circ}$ ) and 4b
($\varphi=90^{\circ}$) illustrate the dependence of $\d E/ \d (\hbar
\omega)\, \d \Omega $ on $\omega$ and on the dimensionless variable
$y=\theta/\theta_0$ with $\theta_0\approx 18\ \mu{\rm rad}$.  It is
seen that in the range of $y=0\dots 0.3$, which corresponds to the
$\theta$-range $\theta=0\dots 5$ $\mu{\rm rad}$, the harmonics are
well resolved.

%%% Fig.4
\begin{figure}
\begin{center}
\hspace{2.5cm}\epsfig{file=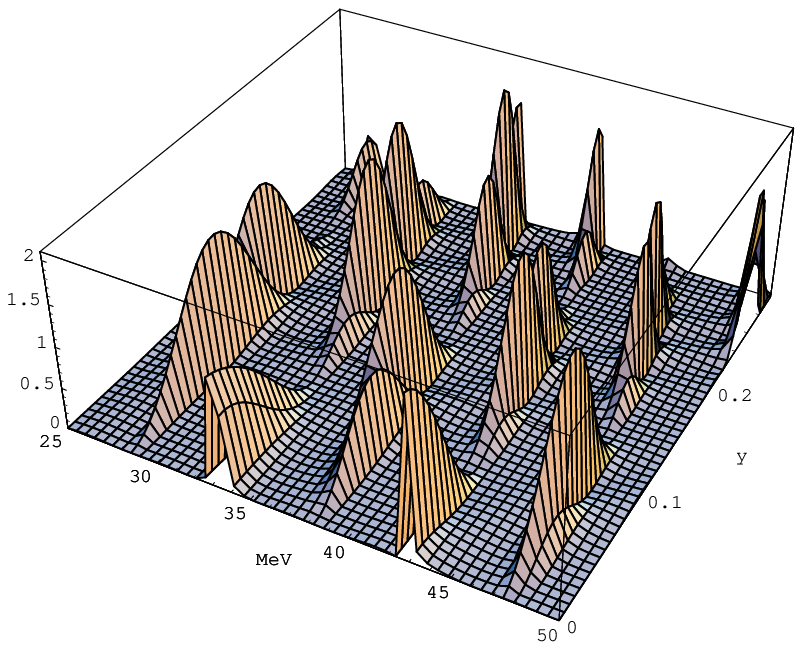,width=9cm}\\
\hspace{2.5cm}\epsfig{file=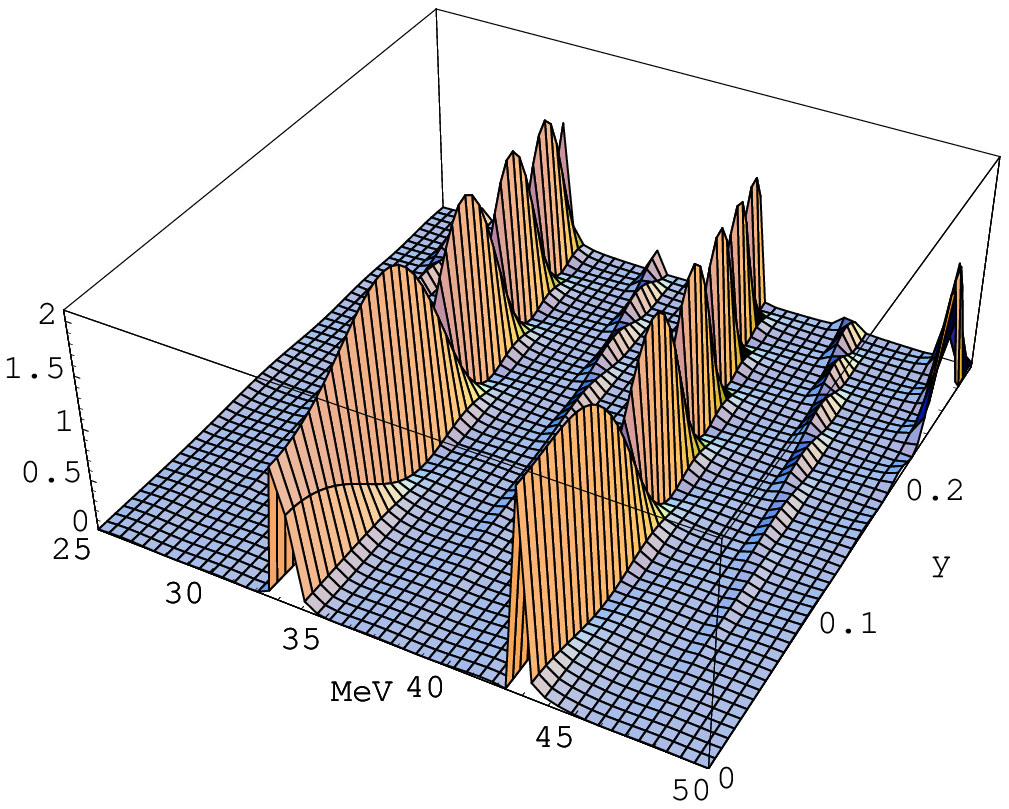,width=9cm}
\end{center}
\caption{The radiated energy (in $10^9$ abs. un.) 
for a 50 GeV positron in carbon crystal versus the photon energy (in
MeV) and the dimensionless variable $y=\theta/\theta_0$ for two values
of the azimuthal angle $\varphi$: (a) $\varphi=0^{\circ}$, (b)
$\varphi=90^{\circ}$.  Other parameters as in figure 3.}
\end{figure}

Two other important characteristics of the radiation, which allow to
discuss the possibility of the stimulated emission, can be obtained
from (\ref{5}).  For simplicity we consider here the case of the
classical limit $\hbar\omega \ll \varepsilon$, and, thus,
$\omega^{\prime}=\omega$ (see (\ref{6c})).

To start let us estimate the quantity $N_{\omega_K}$ 
\begin{equation} 
N_{\omega_K}= 
\left[ 
{\d E \over \hbar\, \d\omega_K\, \d \Omega }\right]_{\theta=0}
\, {\Delta \omega \, \Delta\Omega_K \over \omega_K}
\label{8}
\end{equation}
which is the number of photons of the frequency within the interval
$\omega_K \pm \Delta \omega$ emitted by a particle of the beam in the
cone $\Delta\Omega_K$ (see below) with the axis along the
$z$-direction.  For $\theta=0$ and $\omega_K < \omega_{\rm max}$ the
following expression can be derived from (\ref{5}):
\begin{equation}
\left[
{\d E \over \hbar\, \d \omega_K \, \d \Omega }\right]_{\theta=0}
\approx {\alpha \, N_{\rm u}^2 \, (2K)^{2/3} \over 2\, \theta_0^2}\,
D_N(\eta)
\label{9}
\end{equation}
Making use of (\ref{6a}) one gets
\begin{equation}
\Delta \Omega_K = {2\pi (1 + 2\, p^{-2}) \over K}\, \theta_0^2
\label{10}
\end{equation}
defining the maximal solid angle inside which the $K$th
harmonics is well resolved. Finally, the use of (\ref{7}) for getting
$\Delta \omega$ yields:
\begin{equation} 
N_{\omega_K} \approx 8\pi\, \alpha\, N_{\rm u}\, (2K)^{-4/3}
\label{11}
\end{equation}

Another quantity to be estimated is the gain, $g_K$, which defines the
increase (the case $g_K>0$) or the decrease (if $g_K<0$) per 1 cm in
the total number, $N_K$, of the emitted photons of the frequency
$\omega_K$, $\d N_K = g_K\, N_K\,\d z$, due to the stimulated emission
(absorption) by the particles in the beam.  For the nearly forward
emission this quantity is given by
\begin{equation} 
g_K = 
-(2\pi)^3\, {c^2 \over \omega_K^2}\, \rho\,
{\d \over \d\varepsilon }
\left[
{\d E \over \d \omega_K\, \d \Omega }\right]_{\theta=0}
\, \Delta \omega \, \Delta\Omega_K 
\label{12}
\end{equation}
Here the quantity $\rho$ stands for the density of the beam passing
through the bent crystal.

Being interested in the case $g_K>0$ and noting, that the only
quantity in (\ref{9}) dependent on $\varepsilon$ is the function
$D_N(\eta)$ (see (\ref{6a})), we get the following expression after
carrying out the derivative:
\begin{equation} 
g_K\ (\, {\rm cm}^{-1}\, ) = 
2\pi^2\, \lambda_{\rm c}\,{N_{\omega_K}\, N_{\rm u}\, \lambda 
\over \gamma^3}\, \rho
=
{2 (2\pi)^3 \over (2K)^{4/3} }\, r_{\rm cl}\, 
{ N_{\rm u}^2\, \lambda \over  \gamma^3}\, \rho
\label{13}
\end{equation}
where $\lambda_{\rm c}=\hbar/mc$ is the Compton wavelength and $r_{\rm
cl}= {e^2 /mc^2}$ is the classical radius of a particle ($\lambda_{\rm
c}= 3.9\cdot10^{-11}$ cm, $r_{\rm cl}=2.8\cdot10^{-13}$ cm for a
positron).  The quantities $\lambda$ and $\rho$ are measured in cm and
cm$^{-3}$, respectively.  Note the strong inverse dependence of $g_K$
on $\gamma$ which is due to the radiative recoil. The gain is
proportional to the factor $N_{\rm u}^2$, which reflects the coherence
effect of radiation. The proportionality of the gain to $\lambda$
means that the increase in $\lambda$ leads to the enchancement of the
radiation intensity in the forward direction.

Formulas (\ref{11}) and (\ref{13}) allow to make quantitative
estimates concerning the possibility to obtain the stimulated emission
of the high-energy photons.

Consider the case of the positron beam channeled in a carbon crystal.
It is seen from figure 2, that the number of the undulator periods per
1 cm, $N$, can be easily varied in the range $N = 1\dots 100$ by
tuning the AW parameters.  Therefore, the number of the photons (per
particle) corresponding to the first harmonic, $K=1$, with the energy
$\hbar\omega_1$, and emitted in the crystal of the thickness $L$ is
varied within the interval: $N_{\omega_1}=0.07\cdot(1\dots 100)\cdot
L$, and can be made $\gg 1$. Note that $N_{\omega_1}$ is independent
on $\gamma$.

For numerical estimations we chose $N=20$ cm$^{-1}$, which corresponds
to the AW frequency $\nu=23.3$ MHz. Other parameters used are: $L=10$
cm, $p^2=10$.  The calculated number of the photons is equal to
$N_{\omega_1} = 14$.  To obtain the gain $g_1=1$ cm$^{-1}$ (and, thus,
the total gain $G_1\equiv g_1\, L =10$) at different energies
$\varepsilon$ it is necessary to use the $\rho$-values given in table
1, calculated from (\ref{13}).
\begin{table}
\caption{The magnitudes of the projectile energies, the relativistic
factors, the first harmonic energies and their widths, the AW
amplitudes and the beam densities corresponding to the gain $g_1=1$
cm$^{-1}$ ($G_1\equiv g_1\, L =10$)}
\begin{indented}
\item[]\begin{tabular}{@{}llllll}
\br
$\varepsilon$ & $\gamma$ & $\hbar\omega_1$ & $\Delta \hbar\omega_1$ & 
$a$  & $\rho$  \\
 GeV         &          &                 &                        &
nm & cm$^{-3}$ \\
\br
50 & 10$^5$ & 8.25 MeV & 0.83 MeV & 2.5 & 10$^{22}$ \\
5  & 10$^4$ & 82.5 keV & 8.25 keV & 25  & 10$^{19}$ \\
0.5& 10$^3$ & 0.83 keV & 82.5 eV  & 250 & 10$^{16}$ \\
\br
\end{tabular}
\end{indented}
\end{table}  

It is seen from table 1 that the estimated $\rho$-values strongly
depend on the relativistic factor. Note that even the lagest value of
$\rho$, $10^{22}$ particles in cm$^3$, is comparable with that planned
to achieve within the TESLA project \cite{TESLA}.

Our investigation shows that the described phenomenon can be used for
the construction of the undulator with variable parameters for the
generation of high energy photons.  Also it is shown, that it is
meaningful to discuss the possibility to create a powerful source of
stimulated monochromatic radiation in the keV and the MeV regions.  A
more accurate treatment, which is presently carried out, must take
into account the effects neglected in the present study.  These are:
the channeling radiation (which accompanies the passage of the
channeled particle), the dechanneling processes (which lead to the
decrease in the beam intensity and thus influence the yield of the
photons), the decrease of the emitted photons flux due to the
processes of diffraction and pair-creation, the non-monochromaticity
of the AW, the temperature of the particles in the beam, the
transverse energy distribution of the beam particles and, as a result,
the distribution of the particles in the initial entering angle.

We express our gratitute to Professor H. Klein for helpful discussion.
The authors acknowledge support from the Deutsche
Forschungsgemeinschaft, GSI and BMBF.

\end{document}